# Data Sharing in the PRIMED Consortium: Design, implementation, and recommendations for future policymaking

Johanna L Smith[#,1], Quenna Wong[2], Whitney Hornsby[3], Matthew P Conomos[2], Benjamin D Heavner[2], Iftikhar J Kullo[1], Bruce M Psaty[4], Stephen S Rich[5], Bamidele Tayo[6], Pradeep Natarajan[3], Sarah C Nelson[#,2], Polygenic Risk Methods in Diverse Populations (PRIMED) Consortium Data Sharing Working Group, Polygenic Risk Methods in Diverse Populations (PRIMED) Consortium

[1]Cardiovascular Medicine, Mayo Clinic, Rochester, MN, 55902, USA
[2]Biostatistics, University of Washington, Seattle, WA, 98195, USA
[3]Center for Genomic Medicine and Cardiovascular Research Center, Massachusetts General Hospital, Boston, MA, 02446, USA
[4]Medicine & Epidemiology, University of Washington, Seattle, WA, 98195, USA
[5]Genome Sciences, University of Virginia, Charlottesville, VA, 22908, USA
[6]Public Health Sciences, Loyola University Chicago, Maywood, IL, 60153, USA

[#]*Corresponding Authors*

*Address for correspondence*
Sarah C Nelson, 4333 Brooklyn Ave NE, Seattle, WA, 98105, 206-543-1850, X handle: blueyedgenes, sarahcn@uw.edu

Johanna L Smith, 150 3rd St. SW, Rochester, MN, 55902, 507-284-5647, X handle: Johanna_L_Smith, smith.johanna@mayo.edu

## Abstract

Sharing diverse genomic and other biomedical datasets is critical to advance scientific discoveries and their equitable translation to improve human health. However, data sharing remains challenging in the context of legacy datasets, evolving policies, multi-institutional consortium science, and international stakeholders. The NIH-funded Polygenic Risk Methods in Diverse Populations (PRIMED) Consortium was established to improve the performance of polygenic risk estimates for a broad range of health and disease outcomes with global impacts. Improving polygenic risk score performance across genetically diverse populations requires access to large, diverse cohorts. We report on the design and implementation of data sharing policies and procedures developed in PRIMED to aggregate and analyze data from multiple, heterogeneous sources while adhering to existing data sharing policies for each integrated dataset. We describe two primary data sharing mechanisms: coordinated dbGaP applications and a Consortium Data Sharing Agreement, as well as provide alternatives when individual-level data cannot be shared within the Consortium (e.g., federated analyses). We also describe technical implementation of Consortium data sharing in the NHGRI Analysis Visualization and Informatics Lab-space (AnVIL) cloud platform, to share derived individual-level data, genomic summary results, and methods workflows with appropriate permissions. As a Consortium making secondary use of pre-existing data sources, we also discuss challenges and propose solutions for release of individual- and summary-level data products to the broader scientific community. We make recommendations for ongoing and future policymaking with the goal of informing future consortia and other research activities.

## Introduction

Sharing diverse genomic and other biomedical datasets is critical to advance scientific discoveries and their equitable translation to improve human health. However, data sharing remains challenging in the context of legacy datasets, evolving policies, multi-institutional consortium science, and international stakeholders. The NIH-funded Polygenic Risk Methods in Diverse Populations (PRIMED) Consortium was formed to assess and improve polygenic risk estimates for a broad range of health and disease outcomes with global impacts, with a focus on addressing inequitable performance in diverse populations.[1,2] Comprising seven multi-institutional Study Sites, a Coordinating Center (CC), NIH program staff, and other affiliates and partner programs, the PRIMED Consortium was tasked with identifying and aggregating available geographically, genetically, and ancestrally diverse datasets to improve risk prediction by polygenic risk score (PRS) development. To bridge the performance gap of PRS in diverse groups, the PRIMED Consortium aims to incorporate new and existing genome-wide association study (GWAS) results and leverage methodologic and computational advances by integrating extant genotype and phenotype datasets.

Robust and flexible data sharing policies and procedures are integral to achieving the goals of the PRIMED Consortium. Specifically, increasing overall performance of PRS in diverse groups relies on data availability and the ability to share those data within the Consortium. Subsequent release of derived data products and results to the broader scientific community will facilitate clinical adoption of PRS and further PRS methods development. At the outset of PRIMED, Sites collectively identified approximately 130 data sources of interest, spanning longitudinal cohorts, hospital-based biobanks, and other consortia. These data reside across different public and private repositories with varying mechanisms of access, including the NCBI database of Genotypes and Phenotypes (dbGaP)[3], NIH-designated cloud-based or local repositories (e.g., BioData Catalyst[4]), public non-NIH repositories (e.g., the European Genome-Phenome Archive[5]), and project-specific platforms (e.g., for UK Biobank[6], Million Veteran Program[7], Genes & Health[8], and the All of Us Research Program[9]).

Here we report on the design and implementation of data sharing policies and procedures in the PRIMED Consortium. We build on work of prior consortia and programs[10–13] in developing multiple intra-consortium data sharing mechanisms, implementing cloud-based processes for sharing and collaboration, and maximizing release of derived data and other products to the broader scientific community in alignment with the NIH Genomic Data Sharing and Data Management and Sharing policies (see Web Resources). PRIMED leverages the NHGRI Analysis Visualization and Informatics Lab-space (AnVIL) cloud platform[14] to enable secondary use of heterogeneous data from a variety of sources and locations, and for collaborative data aggregation, harmonization, and analysis. In contrast to consortia generating new primary data, PRIMED's focus on making secondary use of numerous pre-existing datasets requires navigating and generating solutions to different policy challenges. We therefore make recommendations for ongoing and future policymaking with the goal of informing future consortia and other research activities (see Box 1 for summary of recommendations).

## Overview of PRIMED Data Sharing Policies and Agreements

The PRIMED Data Sharing Working Group developed a Data Sharing Policy and related procedures for accessing and sharing heterogeneous and diverse datasets within the Consortium and for releasing Consortium-generated data products to the broader scientific community (see Web Resources). The Data Sharing Policy outlines two primary mechanisms for intra-Consortium data sharing: (1) a Consortium Data Sharing Agreement (CDSA) and (2) coordinated dbGaP applications for data accessible via dbGaP. These mechanisms each establish data sharing circles, composed of investigators with permission to access the same data, such that the data can be shared among them collectively on the AnVIL cloud platform via Consortium shared data workspaces (see "Implementing data sharing in the cloud"). Establishing multiple mechanisms provided flexibility for PRIMED members to contribute data expeditiously through whichever path was most feasible.

### Consortium Data Sharing Agreement

A Consortium Data Sharing Agreement (CDSA) establishes consistent terms and definitions to permit data sharing among all signing groups, and it has emerged as a streamlined, preferable alternative to the set of all pairwise data sharing and transfer agreements in a consortium context.[12,15] We built on the CDSA examples and experiences of prior consortia such as CHARGE[10,12] and C4R[11] to develop a PRIMED-specific CDSA (see Web Resources), with key features including definitions and roles for signing investigators and institutions, distinctions between data contributors (i.e. Data Affiliates) and users, communication of consent and data use limitations, specification of data to be shared (e.g., individual- and/or summary-level data, specific traits) and integration into Consortium governance and other policies (e.g., publications, affiliate membership).

We considered implementation of several strategies adapted from prior consortia[11,12] to minimize the time and effort to develop and implement the PRIMED CDSA, summarized in Box 2. Several tips involve inclusion of a specific clause increasing future administrative flexibility and thereby reducing the frequency at which participating groups obtain institutional signatures for modified documents. Further, we opted to separate CDSA terms from management and implementation aspects that may require updating over time. We also summarize suggestions aimed at streamlining review and approval from participating institutions, as the time to negotiate terms scales with the number of institutions performing legal review (who may have divergent or conflicting modification requirements).

### Coordinated dbGaP applications

PRIMED opted to coordinate dbGaP applications among its investigators to establish data sharing circles for study data accessible via dbGaP. In this model, a set of investigators each submits their own dbGaP application with coordinated elements and parallel Data Access Requests (DARs) to specific studies (see Web Resources). This approach allows sharing data across applications and institutions, both the source data in dbGaP and any cleaned, harmonized, analysis-ready, or otherwise derived versions of the source data. Without the coordination of dbGaP applications, investigators would otherwise not be permitted to share data and derived products across institutions or even with other approved applicants at the same institution.

Paralleling a coordinated application model that other groups have previously adopted,[12] coordinated components of the PRIMED dbGaP application include a common Project title; Research Use Statement; Cloud Use Statement specifying that requested data may be uploaded to and analyzed in the AnVIL cloud platform; and External Collaborators list, which also serves as an Eligibility List specifying individuals permitted to join the data sharing circle—i.e. a Data Access Committee (DAC) will disapprove an application from anyone not on the list. Further, applicants submit DARs for the same or overlapping datasets, ranging from a handful to over a hundred in PRIMED. The development of the procedures relied heavily on review by several DACs that oversee the datasets of interest and were then documented with step-by-step instructions and templates to simplify the application process for investigators, minimize common issues, and streamline the review process for DACs.

### Supplementary sharing mechanisms

In some cases, sharing data within PRIMED using the above data sharing mechanisms may not be feasible. For example, some studies require individual-level data to remain on project-specific research platforms (e.g., the All of Us Researcher Workbench,[9] UK Biobank Research Analysis Platform,[6] Genes & Health Google Cloud Platform Trusted Research Environment,[8] and Million Veteran Program[7]) and/or may be subject to international guidelines, such as the General Data Protection Regulation (GDPR) for cohorts in Europe. Investigators can acquire approved access to some of these platforms, or they can rely on collaboration with other investigators with pre-existing access to perform analyses and share summary-level results. However, the increasing use of trusted research environments that prohibit individual-level data sharing creates an ongoing barrier to pooled analyses across datasets. Similarly, many of the biobanks contributing data for PRIMED analyses have restrictions that prevent them from sharing individual-level data via the PRIMED CDSA. To include these data in Consortium projects, we are also using federated approaches,[16] where analysts affiliated with these biobanks run analyses locally and share allowable summary-level data and/or performance metrics with PRIMED—e.g., into shared Consortium AnVIL workspaces—for further analysis and/or public release.

### Future directions for data sharing mechanisms

Although the above data sharing mechanisms have made possible successful data sharing in multi-institutional collaborations of secondary data users, there are still opportunities for improvement. For example, the coordinated dbGaP application model has not scaled well for collaborations of increasing size (e.g., numbers of investigators and datasets). Applicants must independently manage heterogeneous timing of annual project renewal, any letters of collaboration, and local IRB approval—which, depending on the study and DAC, may also require annual renewal. Across NIH institutes, DACs operate on different timelines and have varying practices when it comes to disapproving requests, with some issuing project-level disapprovals (i.e. all DARs disapproved) when there is an issue with any one DAR. Therefore with coordinated applications, while all of the members of a sharing circle initially have an approved status, over time one or more members may have the status change to disapproved or expired. Ensuring that access to shared consortium data is appropriately managed in this heterogeneous environment adds a significant administrative burden to Consortium research. These factors can lead to disruptions in Consortium scientific progress and, depending on the issues(s), each DAR can take weeks or months to resolve.

As the dbGaP application system is used for data access across multiple NIH repositories,[17] we propose a new type of controlled access data application (e.g., dbGaP, DUOS) intended for multi-institutional collaborations such as a consortium with joint research aims to reduce or eliminate the difficulties of application synchronization, differential expiration timing, and heterogeneous dataset approvals. A single, consortium-level coordinated application that is shared across research institutions could be particularly attractive in the case where data reside in a secure NIH controlled platform that allows data storage and analysis, thereby minimizing the number of project applications requiring submission, maintenance, and review. This would simplify procedures for applicants and DACs, resulting in fewer disruptions to ongoing projects throughout the lifespan of a consortium. We also suggest the creation and use of additional dbGaP Collections, which are aggregates of existing study accessions that can be organized around common consent, scientific project (e.g. consortium or data challenge), phenotype, or other relevant features (see Web Resources). dbGaP Collections mitigate the burden of requesting and maintaining hundreds of DARs per project application by minimizing the number of DARs. We also support efforts to standardize and streamline the operations and review procedures across DACs.[18]

Furthermore, from our experiences across numerous consortia, we have found that early alignment of data sharing expectations (with respect to all data sharing mechanisms) can greatly drive efficiency for establishing sufficient data sharing within and beyond the consortium. Communication can be via funding announcements and early consortium policies and documents that establish terms of award (e.g. conditions, timelines) among funders, awardees, study stakeholders, and collaborators. A terms of consortium participation and award document can indicate broad sharing expectations by specifying acceptable data use limitations, consent, and timeline for key milestones (e.g., dbGaP registration, Institutional Certification, biosample delivery, CDSA concept approvals, data upload/delivery by data type). This approach complements existing NIH data sharing policies and requirements including the Genomic Data Sharing Policy and newer Data Management and Sharing Policy (see Web Resources). Further, whether at the NIH-level or consortium-level, for a consortium with secondary data users such as PRIMED, it will be useful to establish participation and funding contingent on having mechanisms in place for bringing together secondary data sources.

## Implementing data sharing in the cloud

Data sharing within PRIMED is primarily facilitated through the NHGRI Analysis Visualization and Informatics Lab-space (AnVIL) cloud platform,[14] which enables intra-consortium data sharing and collaborative analysis across institutions, with the ability to develop, share, and test new PRS methods and workflows through the use of open source code repositories/registries such as GitHub and Dockstore. PRIMED is one of the first consortia to pilot using AnVIL as secondary users of access-controlled biomedical data.

The Data Sharing Policy and mechanisms described above are operationalized through the organization and management of AnVIL workspaces (see Figure 1). Inside AnVIL workspaces, researchers can run analyses, launch interactive tools like RStudio, Jupyter, and Galaxy, store data, and share results. The CC creates shared Consortium data workspaces on AnVIL to store data provided by the Study Sites or Data Affiliates and manages user access for each workspace, enabling sharing among the appropriate Consortium members. Data contributed via the CDSA is organized such that each data workspace corresponds to a single study-consent group, based on information provided by the contributing Data Affiliate. The CDSA establishes a single data sharing circle, and access is granted to all CDSA data workspaces collectively. Data obtained via coordinated dbGaP application is organized such that each data workspace corresponds to a single version of a study-consent group, based on the study accession number (i.e. phs000000.vX.pY.cZ). Only data from, or derived from, that study accession or a substudy accession where data access is managed by applying to the parent study, is permitted to be stored in that data workspace. Access to dbGaP data workspaces is granted on a per-workspace basis, using the approved DARs from each researcher's coordinated dbGaP application, in effect creating multiple data sharing circles, one for each combination of study, consent group, and version.

Due to the complexity of the PRIMED data sharing policies and scale of the components, the CC built a web application, the AnVIL Consortium Manager (ACM, see Web Resources), to track and manage all of the studies, workspaces, agreements, applications, and members. ACM programmatically interacts with the AnVIL platform via the AnVIL APIs. The CC creates all shared data workspaces using ACM so they can be managed within the application. The CC tracks all signed and executed CDSAs in ACM, and the CDSA representative (usually the Site PI) provides a list of additional members at their signing institution who should have data access covered by their signed and executed CDSA. The CC also uses ACM to maintain a list of submitted coordinated dbGaP applications and periodically checks dbGaP-provided JSON records for up-to-date DAR status at the study-version-consent level for those applications. The application PI provides a list of additional members at their institution (i.e. Internal Collaborators and authorized downloaders on their application, as well as any additional unnamed trainees, lab staff, or postdocs) who should have data access covered by their approved coordinated dbGaP application. PRIMED members register in ACM to authenticate the identity of their AnVIL user account, and the CC adds user accounts to managed groups that correspond to CDSAs and dbGaP applications. The CC grants managed user groups access to data by adding them to the Authorization Domains of the appropriate shared data workspaces. All relationships between user accounts, managed groups, workspaces, agreements, and applications within ACM and between ACM and AnVIL records are audited regularly to ensure that data access is provided only as intended.

## AnVIL utilization for Consortium-wide Studies

The AnVIL platform and the shared data within provides a unique environment to enable collaborative analysis across institutions and implement Consortium-wide projects. Data and metadata shared in the Consortium data workspaces are harmonized and formatted according to the PRIMED common data model[2] (see also Web Resources), which facilitates easily merging data for cross-study analysis. A group of investigators, each with approved access to a common set of shared data workspaces, can establish a shared analysis workspace in AnVIL in which they are permitted to read, merge, and collaboratively analyze data from those data workspaces across studies (Figure 1).

One such Consortium-wide project is benchmarking and comparing the performance of existing PRS methods in admixed populations, with a focus on whether and how they utilize local and global genetic ancestry information—initially GAUDI,[19] PRSice-2,[20] and PRS-CSx.[21] PRIMED investigators are developing Docker images and WDL workflows to run these PRS methods that can be deployed on AnVIL and will be shared publicly via code repositories such as Dockstore and GitHub (see Web Resources). Investigators working on this project have a shared analysis workspace in which they are testing these PRS workflows and applying them to datasets from the shared data workspaces to generate benchmarking statistics that can be used to inform best practices and further development of PRS methods.

Other Consortium-wide efforts are maximizing the current data availability in biobanks affiliated with PRIMED Sites within the limitations of their data sharing policies by utilizing federated analysis approaches, following the efforts of prior consortia.[22,23] One such project is developing new ensemble PRS models that leverage cross-trait information to improve prediction accuracy for priority traits by applying the PRSmix method[24] to existing PRS in the PGS Catalog.[25] During this project, each biobank is harmonizing phenotypes, calculating individual PRS, and performing PRSmix analyses in their chosen computing environment. Summary statistics from the PRSmix analyses for each trait will then be uploaded to a shared analysis workspace in AnVIL for cross-biobank meta-analysis to generate the final combined PRS models. This federated approach allows for biobanks to maintain individual-level data in local computing environments while still enabling cross-Site collaboration and maximizing biobank data utility by aggregating and meta-analyzing shareable summary-level data in AnVIL.

## Release of data to the scientific community

Above we have presented policies and implementation for sharing data within the PRIMED Consortium. Next we discuss sharing (i.e. "release") of data generated by PRIMED to the broader scientific community, ranging from individual-level harmonized genotype and phenotype datasets to summary-level data such as GWAS results and PRS models. We propose future directions for policymakers to facilitate the release of these derived datasets when generated by secondary data users. These suggestions should aid in compliance with the NIH Policy for Data Management and Sharing that went into effect January 2023 (see Web Resources), which aims to maximize the sharing of scientific data from NIH-funded projects.

## Release of individual-level derived data

Building on efforts of multiple consortia,[10,12,26–28] PRIMED has combined and harmonized individual-level phenotype and genotype data from numerous source studies, using the intra-Consortium sharing mechanisms described above to share the source and harmonized data in AnVIL workspaces among Consortium members with requisite access approvals. However, it is not straightforward for a consortium of secondary data users to release these individual-level harmonized data back to the broader scientific community.

Specifically, harmonized phenotypes and imputed genotypes are types of "derivative datasets," defined in the model NIH Data Use Certification Agreement (see Web Resources) as "data derived from controlled-access datasets obtained from NIH-designated data repositories." The Agreement also contains a Non-Transferability section that prohibits the sharing of derived data beyond the approved users on a given dbGaP application, which prevents the further sharing of derived data when the source data are obtained via dbGaP application.

Existing data sharing models require dividing derivative datasets back into corresponding source accessions and working with original study owners/generators to release the harmonized data through those same source accessions.[15] This approach lacks scalability, adds complexity to recent analysis protocol shifts from meta- to pooled analysis approaches that require data reassembly, and is only tractable when the secondary data users can feasibly collaborate/communicate with the original study investigators. Further, dividing data is not possible for some cross-study measures based on combined information, such as pairwise genetic relatedness measures.

Therefore, while we continue to pursue solutions to do so, releasing a combined PRIMED Consortium individual-level dataset may not ultimately be feasible, given the limitations noted above and additional cohort-specific regulations and requirements (i.e. that don't allow sharing individual-level data within the Consortium). Recognizing this, and building on the previous work of the eMERGE Network[29] and TOPMed,[26] we will focus our effort on sharing phenotype harmonization instructions with the scientific community via public repositories such as GitHub or PheKB[30] to enable recreation of harmonized phenotypes from source data.

### Future directions for individual-level derived data sharing

Secondary data users can add substantial value to existing datasets through efforts to harmonize, impute, and aggregate individual-level phenotypic and genotypic data, such as reformatting source data to new data models or translating across data ontologies. However, current barriers to releasing data derived by secondary data users lead to inefficiencies and missed opportunities to both contribute to and benefit from harmonization efforts. While publicly sharing documentation and harmonization algorithms can assist in the recreation of derived datasets, it does not save substantive time, cost, or effort for downstream users and also does not capitalize on the centralized model of data storage and analysis enabled by NIH cloud platforms.

We suggest that policymakers develop policies to facilitate the release of individual-level data derivatives produced by secondary data users, and that data platforms and repositories develop the technical infrastructure to support such release, especially when the source data resides in the given repository. Such policies and procedures need to minimally include: prerequisites for accepting multi-study or single-study derived data, approaches to managing and releasing multi-study derivative datasets, articulation of platform and derived data submitter responsibilities (with respect to quality control, documentation, secondary data submitters, etc.), procedures to notify study stakeholders, and methods to track data provenance and data use limitations.

Relatedly, repositories should also develop robust and flexible ways to track participant consent from source data to individual-level derived data products. Most studies include a process for participants to withdraw consent at any time. Furthermore, especially in longitudinal studies, participants may change consent (e.g., from general research to disease-specific uses) at future study visits or exams. A consent designation for an entire study might also change, e.g., due to administrative changes in the study registration. Such consent updates necessitate repositories hosting derived data products to establish and maintain linkages to the most up-to-date consent information associated with the source data to ensure that repositories are correctly granting access to derived datasets (at the study-consent level) and that downstream uses align with what participants have agreed to. Tracking data provenance and consent designation for individual-level derived data would be further aided by repositories moving towards tracking and communicating (i.e. between repositories) data permissions at the level of individual research participants as opposed to study-consent groups. Note in this scenario it would be important to retain original study context and history. We also suggest repositories develop ways to ingest and release multi-study, individual-level derived datasets in their pooled form, versus split apart into component studies as is current practice (see prior section). Developing technical solutions in repositories to store individual-level data in databases with row-level permissions (as opposed to storing consent group-level data files) would support both data access at the individual-level as well as release of pooled, harmonized datasets. Furthermore, the dbGaP Collections model could be extended to help make pooled, harmonized datasets findable and available to the scientific community.

As an example of progress towards the above recommendations, the TOPMed program developed a harmonized data submission format for data pooled across multiple studies. Specifically, in addition to documentation and algorithms/code, the submission provided individual-level harmonized data provenance (e.g. dbGaP study accession and version associated with the component source data) and instructions how to map to up-to-date individual-level consent and data use limitations (see Stilp at al.[26] Supplement). TOPMed also contributed to the conceptualization of technical infrastructure to support the release of consortium data products (i.e. harmonized phenotypes) as multi-study datasets which can be subsetted to a user's approved study-consent groups.

We contend that early alignment of study protocols with NIH data sharing policies and procedures prior to sample collection for new cohorts can greatly drive efficiency for establishing sufficient data sharing with the scientific community. OurHealth is a new cohort investigating

South Asian-specific genomic and non-genomic risk factors for cardiovascular disease (see Web Resources). OurHealth has a unified, broad consent for general research use (GRU). Institutional Certification was obtained early on, confirming that the submission of large-scale human genomic data generated from OurHealth to an NIH-designated data repository is consistent with the NIH Genomic Data Sharing Policy, the informed consent of the study participants, and research use limitations. PRIMED is funding blended genome exome sequencing (BGE) for 1,000 participants in the OurHealth study. Individual-level genomic and non-genomic data from OurHealth participants will be released on AnVIL, with data access managed through a single dbGaP study accession. Looking ahead, investigators must design cohort studies generating genetic data such as OurHealth with data sharing policies in mind, enabling secondary users to maximize the value of the collected information, but also to facilitate the release of data-derived products, such as genomic summary results, making them available to the broader scientific community.

## Release of summary-level derived data

Sharing summary-level data within consortia and across the broader scientific community has emerged as a practical alternative given the policy and logistical challenges of sharing individual-level genotype and phenotype data (e.g., see above section).[10] Indeed, most methods to *develop* new PRS models only require GWAS summary statistics (e.g., LDpred,[31] lassosum,[32] PRS-CSx,[21] CT-SLEB[33]). While individual-level data are typically required to *tune parameters* and *test* new PRS models, newer methods are being developed that only require summary-level data to perform these additional steps (e.g., PUMAS,[34] SummaryAUC,[35] and PRStuning[36]). Notably, PRS models, i.e. lists of genetic variants with weights, are also summary-level information. In PRIMED, we aim to share summary-level data as broadly as possible, including in open access resources such as the GWAS Catalog[37] and PGS Catalog.[25]

However, in our early Working Group discussions, we found that "summary-level data" was not consistently defined nor understood, which led to ambiguity in determining which existing data sharing policies applied. Specifically, the 2018 Update to NIH Management of Genomic Summary Results (GSR) Access policy (see Web Resources) states that GSR, *"...previously referred to as 'aggregate genomic data' or 'genomic summary statistics,' are generated from primary analyses of genomic research. They convey information relevant to genomic associations with traits or diseases across datasets rather than associations specific to any one individual research participant."* The policy further defines GSR to *"include systematically computed statistics such as, but not limited to: 1) frequency information (e.g., genotype counts and frequencies, or allele counts and frequencies); and 2) association information (e.g., effect size estimates and standard errors, and p-values)."* Notably, the 2018 GSR policy update established a default of unrestricted (i.e. open) access, with the option for submitting institutions to designate a study as "sensitive" and require GSR to be shared through controlled access. Per NIH Points to Consider (see Web Resources), potential study contexts in which the "sensitive" designation may be appropriate include recruitment from a limited geographical area, inclusion of populations or communities at heightened risk of stigmatization and/or discrimination, and/or with rare traits.

The NHGRI Data Sharing Governance Committee determined that PRS scoring files (as defined in Box 3) developed in PRIMED meet the definition of GSR and are therefore under the scope of the NIH GSR policy. While not explicit in the policy, the Committee further clarified that for multi-study GSR (i.e. where multiple studies are used to generate the summary result), the inclusion of one or more "sensitive" studies requires the multi-study GSR to also be shared as controlled access, with the most restrictive data use limitation among the sensitive, contributing studies. These policy determinations had critical implications for PRIMED: scoring files for models developed using one or more sensitive studies can not be fully shared through the open access PGS Catalog, whose submission criteria include sharing "variant information necessary to apply the PGS to new samples (variant rsID and/or genomic position, weights/effect sizes, effect allele, genome build)".[25] In response to this policy determination, we established key PRS terms and definitions in our Consortium Data Sharing Policy and identified which data products fell under the GSR Policy (see Box 3). To further comply with the NIH GSR Policy, we developed programmatic approaches in ACM to track the "sensitive" designation of studies brought into Consortium data sharing circles either through dbGaP applications—where "sensitive" designation is indicated on the Institutional Certification and tracked on the dbGaP study page, or through the CDSA—where data contributors provide the necessary information. To date among the studies with data in AnVIL, 11 of 17 studies shared through our coordinated dbGaP applications and 1 of 3 CDSA studies are designated "sensitive."

We are ultimately trying to navigate complying with the NIH GSR policy while also releasing PRS models as broadly as possible, as championed by the fully open access PGS Catalog.[25] Indeed, PRIMED investigators have made substantial contributions to the PGS Catalog, which the CC tracks by PubMed ID on the Consortium website (see Web Resources; to date 27 publications contributing >200 PRS development and/or evaluation records). For PRS models that require controlled access, one option being considered is to release them through the AnVIL cloud platform, through a new dbGaP study accession established by the CC. Similar dbGaP accessions have previously been established by the CHARGE Consortium (phs000930), the NHLBI TOPMed program (phs001974), and the Million Veteran Program (MVP, phs001672).

### Future directions for summary-level derived data sharing

As a Consortium of secondary data users, we recognize the importance of respectful data stewardship that balances the benefits of open sharing with the potential risks and concerns to study participants and communities, as well as the technical challenges relating to respecting participant consent with respect to derived data. The evolution of NIH policies on GSR sharing reflects this balance as well, with GSR initially shared open access in 2007, moved to restricted access after demonstration of re-identification risks in 2008,[38] and in 2018 iterating to the current state of defaulting to open access GSR sharing while still allowing a path for specific studies to request enhanced protection via controlled access. Based on our experience in PRIMED, we have the following suggestions for NIH and other policymakers to consider for GSR sharing policies and practices moving forward.

Policymakers should distinguish between privacy-based versus respect-based risks when weighing risks and benefits of different GSR sharing approaches. Re-identification is a privacy-based risk,[39] and to date has been the main focus of empirical studies to inform GSR sharing. Alternatively, respect-based risks include the potential for individual- or group-level stigma or data otherwise being used in a way that is objectionable or harmful to the participants (see Nelson et al.[40] for a brief review). Respect-based risks are unlikely to be affected by protections against re-identification (e.g., truncating the number of variants shared from a GWAS), and merit further empirical bioethics research to understand impact and potential mitigation.

Policymakers should ensure a diversity of viewpoints are included in GSR policy-making. Policies that affect the use and sharing of diverse, international datasets—as we are leveraging in PRIMED—should be shaped by a similarly diverse range of perspectives. In highlighting GSR sharing concerns specific to African populations, Tiffin[41] critiqued the NIH GSR Policy development for lack of input from African scientists and communities. Conversely, it is incumbent on researchers and institutions to submit public comments on proposed policies, as stewards of the study populations they work with.

Policymakers should refine and "future proof" the definition of GSR for broader awareness and application of NIH GSR policy moving forward. Apart from the clear examples of allele frequencies, GWAS results, and now PRS models, determining what other analytical products are within scope of the NIH GSR policy requires a clearer and more future-looking definition of GSR. Given broad sharing and reuse of genomic datasets, describing GSR as "generated from primary analyses" and "convey[ing] information relevant to genomic associations with traits or diseases across datasets rather than associations specific to any one individual" (see Web Resources) are likely insufficient to determine if common products such as genetic ancestry models (e.g. SNP loadings for genetic ancestry principal components) are considered GSR. In PRIMED, we had the benefit of dialogue with NIH program staff and policy experts about applicability of the GSR policy in our specific Consortium context, but individual researchers and smaller groups are unlikely to have awareness, bandwidth, and access to delve into such policy deliberations and thus may unintentionally share GSR inappropriately.

Policymakers should revisit the risk-benefit analysis for multi-study GSR sharing and clarify in NIH policy when and why multi-study GSR require controlled access sharing. The NIH GSR Policy does not explicitly address how multi-study GSR involving one or more "sensitive" studies should be shared, though combining individual studies for further discovery (e.g. meta- or pooled-analysis) is an increasingly common use case in large-scale genomic research and almost an imperative in PRS model development. PRIMED was informed that for multi-study GSR, the inclusion of one or more "sensitive" studies requires the multi-study GSR to also be controlled access, based in part on precedent for registering multi-study dbGaP accessions. However, anecdotally we have heard from some PIs of "sensitive" studies that they are not concerned with fully open access GSR sharing if the study is combined with other datasets. The current Institutional Certification process does not capture this nuance of single- vs multi-study GSR sharing and therefore may unnecessarily push multi-study GSR into controlled access with a narrow consent designation. As in the single-study GSR context, a risk-benefit analysis that accounts for both privacy- and respect-based risks to the contributing study populations should guide how multi-study GSR are shared.

## Conclusion

The PRIMED Consortium has implemented multiple methods to share data and products within the Consortium and with the broader scientific community, supporting methods development to improve risk prediction in diverse, global populations. Our data sharing efforts are situated within a larger genomic data sharing landscape[e.g.,42,43] and provide an example of some of the tensions created by aiming to broaden data sharing in the context of legacy datasets, evolving policies, multi-institutional consortium science, and international stakeholders. Here we reported on how we approached and adapted to challenges of using heterogeneous, pre-existing datasets with a variety of use restrictions, access procedures, and data locations. We have summarized future considerations and suggestions for policy making (Box 1), reflecting approaches that have supported PRIMED collaborative research and are intended to promote efficiency in future programs/consortia. Some suggestions have been implemented in the PRIMED Consortium through prospective cohorts, such as Our Health. Though some of our suggestions require technical advancement and policy innovations, it is our aim to highlight and suggest solutions for barriers in genomic research across diverse populations moving forward.

## Acknowledgements

This publication was supported by the National Institutes of Health for the project "Polygenic Risk Methods in Diverse Populations (PRIMED) Consortium", with grant funding for Study Sites CARDINAL (U01HG011717), CAPE (U01HG011715), D-PRISM (U01HG011723), EPIC-PRS (U01HG011720), FFAIRR-PRS (U01HG011719), PRIMED-Cancer (U01CA261339), PREVENT (U01HG011710), and the Coordinating Center (U01HG011697). The content is solely the responsibility of the authors and does not necessarily represent the official views of the National Institutes of Health. In addition, J.L.S acknowledges HL07111-45, and B.M.P. acknowledges HL105756.

## Conflicts of Interest

P.N. reports research grants from Allelica, Amgen, Apple, Boston Scientific, Genentech / Roche, and Novartis, personal fees from Allelica, Apple, AstraZeneca, Blackstone Life Sciences, Bristol Myers Squibb, Creative Education Concepts, CRISPR Therapeutics, Eli Lilly & Co, Esperion Therapeutics, Foresite Capital, Foresite Labs, Genentech / Roche, GV, HeartFlow, Magnet Biomedicine, Merck, Novartis, Novo Nordisk, TenSixteen Bio, and Tourmaline Bio, equity in Bolt, Candela, Mercury, MyOme, Parameter Health, Preciseli, and TenSixteen Bio, and spousal employment at Vertex Pharmaceuticals, all unrelated to the present work.

# Figures and Tables

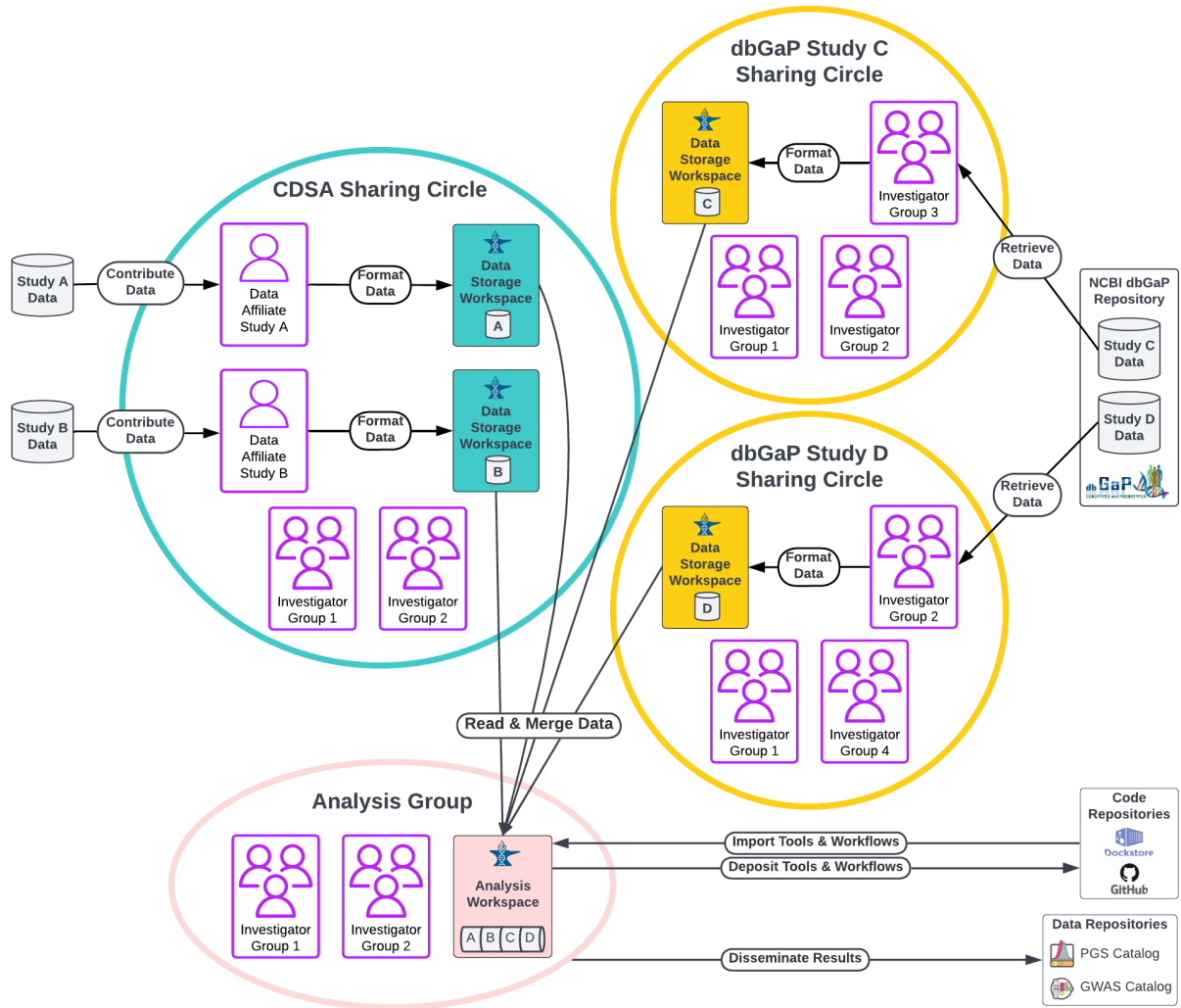

Figure 1. **PRIMED Data Sharing Ecosystem.** Data sharing circles are established by PRIMED data sharing policy. The CDSA establishes one data sharing circle (teal), and data affiliates contribute their study's data to a designated data storage workspace. The coordinated dbGaP applications establish many data sharing circles (gold)—one per study, consent group, version combination—and PRIMED investigators contribute data they retrieve from the dbGaP repository to a designated data storage workspace. Data is formatted to the PRIMED common data model before sharing. The CDSA representative, or dbGaP application PI, provides the PRIMED CC a list of additional investigators who have data access permission covered by their agreement or application, respectively; these investigator groups are granted read access to the appropriate data storage workspaces using the CC's ACM tool. Investigator groups in multiple data sharing circles are permitted to form an analysis group (pink), where they can read and merge study data from the data storage workspaces to which they have common access into a single analysis workspace. Analysis tools and workflows can be imported from code repositories into analysis workspaces. Newly developed tools and workflows are deposited into code repositories, and summary results are deposited to data repositories (when permissible).

Box 1. Summary of recommendations for future directions

List of future-looking recommendations for different audiences (i.e. policymakers, data platforms, research consortia, data generators, and other researchers) across the activities of within-consortium data sharing, release of individual-level derived data, and release of summary-level derived data. Recommendations that span more than one activity are in merged columns.

| Summary of Future Directions | | |
|---|---|---|
| Within-consortium Data Sharing | Release of Individual-level Data | Release of Summary-level Data |
| Policymakers establish a new type of controlled access data application for multi-institutional collaborations with joint research aims (e.g., a single, consortium-level coordinated dbGaP application) | Policymakers develop policies to facilitate the release of individual-level data derivatives produced by secondary data users | Researchers and policymakers distinguish between privacy- and respect-based risks when weighing the risk/benefits of different GSR sharing approaches |
| Platforms (e.g., dbGaP, DUOS) update Data Access Request systems to support such multi-institutional, consortium-level applications | Platforms develop robust and flexible ways to track participant consent in derived datasets | Policymakers ensure diverse and international viewpoints are included in GSR policy-making |
| Funders establish clear and specific terms of award and conditions of consortium participation (e.g., deliverables with timelines) among awardees, study stakeholders, and collaborators | Platforms develop infrastructure to ingest and release multi-study, individual-level derived datasets in their pooled form | Policymakers establish a clear, complete, and future-looking GSR definition for broader awareness and application of NIH GSR policy |
| Consortia can create new consortium-centric dbGaP Collections of datasets across multiple study accessions to make data used, generated, and harmonized together findable, simplify dbGaP application submission and maintenance for both members of the consortium and the broader scientific community, and streamline the DAC review process | | Policymakers revisit the risk-benefit analysis for multi-study GSR sharing and clarify in NIH policy when and why multi-study GSR require controlled access sharing |
| | Data generators align study protocols and informed consent procedures with NIH data sharing policies prior to sample collection for new cohorts, to facilitate downstream release of individual-and summary-level data | |

Box 2. Tips and language for consideration in Consortium Data Sharing Agreements.

Suggested approaches for creating and maintaining Consortium Data Sharing Agreements (CDSA), based on our experience in PRIMED and guidance from other consortia (e.g. CHARGE, C4R).

| | |
|---|---|
| **Administrative Flexibility** | Include data access and use by investigators, fellows, students, and research staff at the same institution as designated by signee |
| | Allow developer data access, for non-research purposes |
| | Specify sharing to be inclusive of derived, harmonized, imputed, or re-processed data |
| | Allow combining of CDSA data with data from non-CDSA sources |
| | Permit release of derived data to the scientific community |
| | Allow migration/extension of data to additional NIH-designated repositories and cloud service providers and platforms |
| | Use of universal (vs U.S.-centric) language to maximize inclusiveness of signees (e.g. consider international guidelines, such as GDPR for European regulations) |
| **Core Functions and Features** | Allow dynamic membership that accommodates new signees to join as they are able to sign (vs requiring initial specification of participating groups) |
| | Define terms specific to types of signees (e.g. data generators, secondary data users) |
| | Specify overseeing committee infrastructure whose charge may include matters of membership, voter representation, and responsibilities |
| **Institutional Review and Approval** | Ask for pre-review where possible (i.e. before routing for signature) |
| | Have legal teams speak with each other directly in the case of conflicting modifications |
| | Identify key datasets, active projects, study stakeholders, and institutions to consult during the legal review process (and for streamlining future amendment processes) |
| | Include options to sign a brief addendum letter summarizing updates (vs signing a new CDSA version) |

Box 3. PRS* Terms and Definitions

Defined terms related to PRS development, validation, and implementation, based on prior literature and Consortium discussions. For ease of identifying applicable data sharing policies, each term is classified by data type.

| PRS Term and Definition | Data Classification |
|---|---|
| **PRS method / algorithm:** analytical/statistical approach to analyze individual- or summary-level data and generate a new PRS model<br>● e.g., LD-pred, PRS-CSx, GAUDI, P&T | *Not data* |
| **PRS model**<br>● **Scoring file:** a list of variants with associated weights/effect sizes<br>   ○ e.g., the data in the PGS Catalog Scoring Files<br>● **Metadata**: additional details about the model<br>   ○ e.g., studies and populations used to develop and/or validate the model; PRS method used to develop the model; number of variants included; genome build | *Genomic Summary Results (GSR)***<br><br>*Summary-level data (not GSR); reported in publications* |
| **Individual PRS:** the output numeric value calculated from a PRS model for an individual. This may be a raw score or may be presented as an adjusted score in the context of a population distribution – i.e. as a percentile or number of standard deviations from the mean.<br>● e.g., raw: 2.1, -1.3; adjusted: 97th percentile | *Individual-level data* |
| **PRS performance metrics:** numerical metrics describing performance of the PRS model in some population(s)<br>● e.g., PRS effect size; $R^2$ (proportion of variance explained); AUC (or other classification metrics) | *Summary-level data (not GSR); reported in publications* |

*\*Terms "PRS," "PGS," and "GRS" (genetic or genomic risk score) can be used interchangeably for these definitions in this policy context (see also Wand et al. 2021 for distinctions); may also be referred to as "PGI" (polygenic index) in some disciplines (Burt et al. 2024)*

*\*\*GSR derived from one or more studies with a sensitive designation can only be shared via controlled access. See Data Sharing outside of PRIMED for further information.*

## Web Resources

NIH Genomic Data Sharing Policy, https://grants.nih.gov/grants/guide/notice-files/NOT-OD-14-124.html

Final NIH Policy for Data Management and Sharing, https://grants.nih.gov/grants/guide/notice-files/NOT-OD-21-013.html

PRIMED Data Sharing Policy, https://primedconsortium.org/about/policies/data-sharing-policy

PRIMED Consortium Data Sharing Agreement, https://primedconsortium.org/about/policies/consortium-data-sharing-agreement

dbGaP FAQs related to Coordinated Applications: https://www.ncbi.nlm.nih.gov/books/NBK570251/#DAreq_Collaborators.i_am_an_authorized_u and https://www.ncbi.nlm.nih.gov/books/NBK570242/#DAreq_ControlledAcc.would_you_give_me_a

dbGaP Collections. https://www.ncbi.nlm.nih.gov/projects/gap/cgi-bin/GetCollectionList.cgi

AnVIL Consortium Manager, https://github.com/UW-GAC/django-anvil-consortium-manager

PRIMED common data model, https://github.com/UW-GAC/primed_data_models

PRIMED Dockstore, PRS collection: https://dockstore.org/organizations/PRIMED/collections/PRS

Model NIH Data Use Certification Agreement, https://sharing.nih.gov/sites/default/files/flmngr/Universal_DUC.pdf

OurHealth study: https://ourhealthstudy.org/

2018 Update to NIH Management of Genomic Summary Results (GSR) Access Policy, https://grants.nih.gov/grants/guide/notice-files/NOT-OD-19-023.html

Points to Consider for Institutions and Institutional Review Boards in Submission and Secondary Use of Human Genomic Data under the National Institutes of Health Genomic Data Sharing Policy, https://sharing.nih.gov/sites/default/files/flmngr/GDS_Points_to_Consider_for_Institutions_and_IRBs.pdf

# References

1. Kachuri, L., Chatterjee, N., Hirbo, J., Schaid, D.J., Martin, I., Kullo, I.J., Kenny, E.E., Pasaniuc, B., PRIMED Methods Working Group, Witte, J.S., et al. Principles and methods for transferring polygenic risk scores across global populations. Nat. Rev. Genet. *(in press)*.
2. Kullo, I.J., Conomos, M.P., Nelson, S.C., Adebamowo, S.N., Choudhury, A., Conti, D., Fullerton, S.M., Gogarten, S.M., Heavner, B., Hornsby, W.E., et al. (2024). The PRIMED Consortium: Reducing disparities in polygenic risk assessment. Am. J. Hum. Genet. *111*, 2594–2606. https://doi.org/10.1016/j.ajhg.2024.10.010.
3. Mailman, M.D., Feolo, M., Jin, Y., Kimura, M., Tryka, K., Bagoutdinov, R., Hao, L., Kiang, A., Paschall, J., Phan, L., et al. (2007). The NCBI dbGaP database of genotypes and phenotypes. Nat. Genet. *39*, 1181–1186. https://doi.org/10.1038/ng1007-1181.
4. Ahalt, S., Avillach, P., Boyles, R., Bradford, K., Cox, S., Davis-Dusenbery, B., Grossman, R.L., Krishnamurthy, A., Manning, A., Paten, B., et al. (2023). Building a collaborative cloud platform to accelerate heart, lung, blood, and sleep research. J. Am. Med. Inform. Assoc. JAMIA *30*, 1293–1300. https://doi.org/10.1093/jamia/ocad048.
5. Lappalainen, I., Almeida-King, J., Kumanduri, V., Senf, A., Spalding, J.D., Ur-Rehman, S., Saunders, G., Kandasamy, J., Caccamo, M., Leinonen, R., et al. (2015). The European Genome-phenome Archive of human data consented for biomedical research. Nat. Genet. *47*, 692–695. https://doi.org/10.1038/ng.3312.
6. Allen, N.E., Lacey, B., Lawlor, D.A., Pell, J.P., Gallacher, J., Smeeth, L., Elliott, P., Matthews, P.M., Lyons, R.A., Whetton, A.D., et al. (2024). Prospective study design and data analysis in UK Biobank. Sci. Transl. Med. *16*, eadf4428. https://doi.org/10.1126/scitranslmed.adf4428.
7. Gaziano, J.M., Concato, J., Brophy, M., Fiore, L., Pyarajan, S., Breeling, J., Whitbourne, S., Deen, J., Shannon, C., Humphries, D., et al. (2016). Million Veteran Program: A mega-biobank to study genetic influences on health and disease. J. Clin. Epidemiol. *70*, 214–223. https://doi.org/10.1016/j.jclinepi.2015.09.016.
8. Finer, S., Martin, H.C., Khan, A., Hunt, K.A., MacLaughlin, B., Ahmed, Z., Ashcroft, R., Durham, C., MacArthur, D.G., McCarthy, M.I., et al. (2020). Cohort Profile: East London Genes & Health (ELGH), a community-based population genomics and health study in British Bangladeshi and British Pakistani people. Int. J. Epidemiol. *49*, 20–21i. https://doi.org/10.1093/ije/dyz174.
9. All of Us Research Program Genomics Investigators (2024). Genomic data in the All of Us Research Program. Nature *627*, 340–346. https://doi.org/10.1038/s41586-023-06957-x.
10. Psaty, B.M., O'Donnell, C.J., Gudnason, V., Lunetta, K.L., Folsom, A.R., Rotter, J.I., Uitterlinden, A.G., Harris, T.B., Witteman, J.C.M., Boerwinkle, E., et al. (2009). Cohorts for Heart and Aging Research in Genomic Epidemiology (CHARGE) Consortium: Design of prospective meta-analyses of genome-wide association studies from 5 cohorts. Circ. Cardiovasc. Genet. *2*, 73–80. https://doi.org/10.1161/CIRCGENETICS.108.829747.
11. Oelsner, E.C., Krishnaswamy, A., Balte, P.P., Allen, N.B., Ali, T., Anugu, P., Andrews, H.F., Arora, K., Asaro, A., Barr, R.G., et al. (2022). Collaborative Cohort of Cohorts for COVID-19 Research (C4R) Study: Study Design. Am. J. Epidemiol. *191*, 1153–1173. https://doi.org/10.1093/aje/kwac032.
12. Brody, J.A., Morrison, A.C., Bis, J.C., O'Connell, J.R., Brown, M.R., Huffman, J.E., Ames, D.C., Carroll, A., Conomos, M.P., Gabriel, S., et al. (2017). Analysis commons, a team approach to discovery in a big-data environment for genetic epidemiology. Nat. Genet. *49*, 1560–1563. https://doi.org/10.1038/ng.3968.
13. Taliun, D., Harris, D.N., Kessler, M.D., Carlson, J., Szpiech, Z.A., Torres, R., Taliun, S.A.G., Corvelo, A., Gogarten, S.M., Kang, H.M., et al. (2021). Sequencing of 53,831 diverse genomes from the NHLBI TOPMed Program. Nature *590*, 290–299.

https://doi.org/10.1038/s41586-021-03205-y.
14. Schatz, M.C., Philippakis, A.A., Afgan, E., Banks, E., Carey, V.J., Carroll, R.J., Culotti, A., Ellrott, K., Goecks, J., Grossman, R.L., et al. (2022). Inverting the model of genomics data sharing with the NHGRI Genomic Data Science Analysis, Visualization, and Informatics Lab-space. Cell Genomics *2*, 100085. https://doi.org/10.1016/j.xgen.2021.100085.
15. Psaty, B.M., Rich, S.S., and Boerwinkle, E. (2019). Innovation in Genomic Data Sharing at the NIH. N. Engl. J. Med. *380*, 2192–2195. https://doi.org/10.1056/NEJMp1902363.
16. Casaletto, J., Bernier, A., McDougall, R., and Cline, M.S. (2023). Federated Analysis for Privacy-Preserving Data Sharing: A Technical and Legal Primer. Annu. Rev. Genomics Hum. Genet. *24*, 347–368. https://doi.org/10.1146/annurev-genom-110122-084756.
17. Dahlquist, J.M., Nelson, S.C., and Fullerton, S.M. (2023). Cloud-based biomedical data storage and analysis for genomic research: Landscape analysis of data governance in emerging NIH-supported platforms. HGG Adv. *4*, 100196. https://doi.org/10.1016/j.xhgg.2023.100196.
18. Lawson, J., Rahimzadeh, V., Baek, J., and Dove, E.S. (2024). Achieving Procedural Parity in Managing Access to Genomic and Related Health Data: A Global Survey of Data Access Committee Members. Biopreservation Biobanking *22*, 123–129. https://doi.org/10.1089/bio.2022.0205.
19. Sun, Q., Rowland, B.T., Chen, J., Mikhaylova, A.V., Avery, C., Peters, U., Lundin, J., Matise, T., Buyske, S., Tao, R., et al. (2024). Improving polygenic risk prediction in admixed populations by explicitly modeling ancestral-differential effects via GAUDI. Nat. Commun. *15*, 1016. https://doi.org/10.1038/s41467-024-45135-z.
20. Choi, S.W., and O'Reilly, P.F. (2019). PRSice-2: Polygenic Risk Score software for biobank-scale data. GigaScience *8*. https://doi.org/10.1093/gigascience/giz082.
21. Ruan, Y., Lin, Y.-F., Feng, Y.-C.A., Chen, C.-Y., Lam, M., Guo, Z., Stanley Global Asia Initiatives, He, L., Sawa, A., Martin, A.R., et al. (2022). Improving polygenic prediction in ancestrally diverse populations. Nat. Genet. *54*, 573–580. https://doi.org/10.1038/s41588-022-01054-7.
22. Nikpay, M., Goel, A., Won, H.-H., Hall, L.M., Willenborg, C., Kanoni, S., Saleheen, D., Kyriakou, T., Nelson, C.P., Hopewell, J.C., et al. (2015). A comprehensive 1,000 Genomes-based genome-wide association meta-analysis of coronary artery disease. Nat. Genet. *47*, 1121–1130. https://doi.org/10.1038/ng.3396.
23. Zhou, W., Kanai, M., Wu, K.-H.H., Rasheed, H., Tsuo, K., Hirbo, J.B., Wang, Y., Bhattacharya, A., Zhao, H., Namba, S., et al. (2022). Global Biobank Meta-analysis Initiative: Powering genetic discovery across human disease. Cell Genomics *2*, 100192. https://doi.org/10.1016/j.xgen.2022.100192.
24. Truong, B., Hull, L.E., Ruan, Y., Huang, Q.Q., Hornsby, W., Martin, H., van Heel, D.A., Wang, Y., Martin, A.R., Lee, S.H., et al. (2024). Integrative polygenic risk score improves the prediction accuracy of complex traits and diseases. Cell Genomics *4*, 100523. https://doi.org/10.1016/j.xgen.2024.100523.
25. Lambert, S.A., Gil, L., Jupp, S., Ritchie, S.C., Xu, Y., Buniello, A., McMahon, A., Abraham, G., Chapman, M., Parkinson, H., et al. (2021). The Polygenic Score Catalog as an open database for reproducibility and systematic evaluation. Nat. Genet. *53*, 420–425. https://doi.org/10.1038/s41588-021-00783-5.
26. Stilp, A.M., Emery, L.S., Broome, J.G., Buth, E.J., Khan, A.T., Laurie, C.A., Wang, F.F., Wong, Q., Chen, D., D'Augustine, C.M., et al. (2021). A System for Phenotype Harmonization in the National Heart, Lung, and Blood Institute Trans-Omics for Precision Medicine (TOPMed) Program. Am. J. Epidemiol. *190*, 1977–1992. https://doi.org/10.1093/aje/kwab115.
27. Musunuru, K., Lettre, G., Young, T., Farlow, D.N., Pirruccello, J.P., Ejebe, K.G., Keating, B.J., Yang, Q., Chen, M.-H., Lapchyk, N., et al. (2010). Candidate gene association resource (CARe): design, methods, and proof of concept. Circ. Cardiovasc. Genet. *3*, 267–275. https://doi.org/10.1161/CIRCGENETICS.109.882696.
28. Oelsner, E.C., Balte, P.P., Cassano, P.A., Couper, D., Enright, P.L., Folsom, A.R., Hankinson, J., Jacobs, D.R., Kalhan, R., Kaplan, R., et al. (2018). Harmonization of Respiratory Data From 9 US Population-Based Cohorts: The NHLBI Pooled Cohorts Study. Am. J. Epidemiol. *187*, 2265–2278. https://doi.org/10.1093/aje/kwy139.
29. Newton, K.M., Peissig, P.L., Kho, A.N., Bielinski, S.J., Berg, R.L., Choudhary, V., Basford, M., Chute, C.G., Kullo, I.J., Li, R., et al. (2013). Validation of electronic medical record-based phenotyping algorithms: results and lessons learned from the eMERGE network. J. Am. Med. Inform. Assoc. JAMIA *20*, e147-154. https://doi.org/10.1136/amiajnl-2012-000896.
30. Kirby, J.C., Speltz, P., Rasmussen, L.V., Basford, M., Gottesman, O., Peissig, P.L., Pacheco, J.A., Tromp, G., Pathak, J., Carrell, D.S., et al. (2016). PheKB: a catalog and workflow for creating electronic phenotype algorithms for transportability. J. Am. Med. Inform. Assoc. JAMIA *23*, 1046–1052. https://doi.org/10.1093/jamia/ocv202.
31. Vilhjálmsson, B.J., Yang, J., Finucane, H.K., Gusev, A., Lindström, S., Ripke, S., Genovese, G., Loh, P.-R., Bhatia, G., Do, R., et al. (2015). Modeling Linkage Disequilibrium Increases Accuracy of Polygenic Risk Scores. Am. J. Hum. Genet. *97*, 576–592. https://doi.org/10.1016/j.ajhg.2015.09.001.
32. Mak, T.S.H., Porsch, R.M., Choi, S.W., Zhou, X., and Sham, P.C. (2017). Polygenic scores via penalized regression on summary statistics. Genet. Epidemiol. *41*, 469–480. https://doi.org/10.1002/gepi.22050.
33. Zhang, H., Zhan, J., Jin, J., Zhang, J., Lu, W., Zhao, R., Ahearn, T.U., Yu, Z., O'Connell, J., Jiang, Y., et al. (2023). A new method for multiancestry polygenic prediction improves performance across diverse populations. Nat. Genet. *55*, 1757–1768. https://doi.org/10.1038/s41588-023-01501-z.
34. Zhao, Z., Yi, Y., Song, J., Wu, Y., Zhong, X., Lin, Y., Hohman, T.J., Fletcher, J., and Lu, Q. (2021). PUMAS: fine-tuning polygenic risk scores with GWAS summary statistics. Genome Biol. *22*, 257. https://doi.org/10.1186/s13059-021-02479-9.
35. Song, L., Liu, A., Shi, J., and Molecular Genetics of Schizophrenia Consortium (2019). SummaryAUC: a tool for evaluating the performance of polygenic risk prediction models in validation datasets with only summary level statistics. Bioinforma. Oxf. Engl. *35*, 4038–4044. https://doi.org/10.1093/bioinformatics/btz176.
36. Jiang, W., Chen, L., Girgenti, M.J., and Zhao, H. (2024). Tuning parameters for polygenic risk score methods using GWAS summary statistics from training data. Nat. Commun. *15*, 24. https://doi.org/10.1038/s41467-023-44009-0.
37. Sollis, E., Mosaku, A., Abid, A., Buniello, A., Cerezo, M., Gil, L., Groza, T., Güneş, O., Hall, P., Hayhurst, J., et al. (2023). The NHGRI-EBI GWAS Catalog: knowledgebase and deposition resource. Nucleic Acids Res. *51*, D977–D985. https://doi.org/10.1093/nar/gkac1010.
38. Homer, N., Szelinger, S., Redman, M., Duggan, D., Tembe, W., Muehling, J., Pearson, J.V., Stephan, D.A., Nelson, S.F., and Craig, D.W. (2008). Resolving individuals contributing trace amounts of DNA to highly complex mixtures using high-density SNP genotyping microarrays. PLoS Genet. *4*, e1000167. https://doi.org/10.1371/journal.pgen.1000167.
39. Bonomi, L., Huang, Y., and Ohno-Machado, L. (2020). Privacy challenges and research opportunities for genomic data sharing. Nat. Genet. *52*, 646–654. https://doi.org/10.1038/s41588-020-0651-0.
40. Nelson, S.C., Gogarten, S.M., Fullerton, S.M., Isasi, C.R., Mitchell, B.D., North, K.E., Rich, S.S., Taylor, M.R.G., Zöllner, S., and Sofer, T. (2022). Social and scientific motivations to move beyond groups in allele frequencies: The TOPMed experience. Am. J. Hum. Genet. *109*, 1582–1590. https://doi.org/10.1016/j.ajhg.2022.07.008.

41. Tiffin, N. (2019). Potential risks and solutions for sharing genome summary data from African populations. BMC Med. Genomics *12*, 152. https://doi.org/10.1186/s12920-019-0604-6.
42. Rehm, H.L., Page, A.J.H., Smith, L., Adams, J.B., Alterovitz, G., Babb, L.J., Barkley, M.P., Baudis, M., Beauvais, M.J.S., Beck, T., et al. (2021). GA4GH: International policies and standards for data sharing across genomic research and healthcare. Cell Genomics *1*, 100029. https://doi.org/10.1016/j.xgen.2021.100029.
43. Stark, Z., Glazer, D., Hofmann, O., Rendon, A., Marshall, C.R., Ginsburg, G.S., Lunt, C., Allen, N., Effingham, M., Hastings Ward, J., et al. (2025). A call to action to scale up research and clinical genomic data sharing. Nat. Rev. Genet. *26*, 141–147. https://doi.org/10.1038/s41576-024-00776-0.